\documentclass[journal]{IEEEtran}
\usepackage{graphicx}

\begin{document}
\title{Test of a prototype neutron spectrometer based on diamond detectors in a fast reactor}
%
%

\author{M.~Osipenko,
	F.~Pompili,
        M.~Ripani,
	M.~Pillon,
	G.~Ricco,
	B.~Caiffi,
	R.~Cardarelli,
	G.~Verona-Rinati,
	and~S.~Argiro
\thanks{Manuscript received April 6, 2015. This work was supported in part by the Istituto Nazionale di Fisica Nucleare INFN-E strategic program.}
\thanks{M.~Osipenko, M.~Ripani and G.~Ricco are with the Istituto Nazionale di Fisica Nucleare, Genoa, 16146 Italy (e-mail: osipenko@ge.infn.it).}%
\thanks{B.~Caiffi, was with Universit\`a di Genova, Genoa, 16146 Italy.}%
\thanks{F.~Pompili, M.~Pillon and M.~Angelone are with the ENEA, Frascati, 00044 Italy.}%
\thanks{G.~Verona-Rinati and R.~Cardarelli are with Universit\`a di Roma Tor Vergata, Rome, 00133 Italy.}%
\thanks{S.~Argiro is with Universit\`a di Torino and INFN, Turin, 10125 Italy.}%

}

\maketitle
\pagestyle{empty}
\thispagestyle{empty}

\begin{abstract}
A prototype of neutron spectrometer based on diamond detectors has been developed.
This prototype consists of a $^6$Li neutron converter sandwiched between two CVD diamond crystals.
The radiation hardness of the diamond crystals makes it suitable for applications in low power research reactors,
while a low sensitivity to gamma rays and low leakage current of the detector permit to reach good energy resolution. A fast coincidence between two crystals is used to reject background.
The detector was read out using two different electronic chains connected to it by a few meters of cable. The first chain was based on conventional charge-sensitive amplifiers, the other used a custom fast charge amplifier developed for this purpose.
The prototype has been tested at various neutron sources and showed its practicability. In particular, the detector was calibrated in a TRIGA thermal reactor (LENA laboratory, University of Pavia) with neutron fluxes of $10^8$ n/cm$^2$s and at the 3 MeV D-D monochromatic neutron source named FNG (ENEA, Rome) with neutron fluxes of $10^6$ n/cm$^2$s. The neutron spectrum measurement was performed at the TAPIRO fast research reactor (ENEA, Casaccia) with fluxes of 10$^9$ n/cm$^2$s. The obtained spectra were compared to Monte Carlo simulations, modeling detector response with MCNP and Geant4.
\end{abstract}

\begin{IEEEkeywords}
neutron spectrometer, fast reactor, diamond detector
\end{IEEEkeywords}

\section{Introduction}
\IEEEPARstart{T}{he} diamond detectors were proposed for measurements of neutrons in nuclear facilities
already in the '60s.
This was motivated by the outstanding properties of diamond, among others its high radiation hardness
and low intrinsic noise at high temperatures.
In fact, the Wigner displacement energy necessary to remove an atom from diamond crystal is 43 eV
compared to 13-20 eV of Silicon. Many studies were undertaken to estimate diamond radiation hardness~\cite{cvd_rad_hard}
and concluded that it is at least an order of magnitude higher than that of Si. The bandgap of diamond is 5.5 eV
allowing only negligible concentration of carriers at room temperature, mostly created by impurities
and crystal defects. All these advantages of diamond crystal were recognized long ago.
However, only with advent of the CVD technique practical
applications of diamond detectors became feasible. In fact, CVD diamond growth provides unprecedented
crystal quality, mandatory for detector applications.
In particular, the relative concentration of most common diamond impurities like B and N was reduced
below $(1\div 5)\times 10^{-9}$~\cite{cvd_e6}. Also a high quality polishing techniques were developed
providing nm scale surface uniformity~\cite{cvd_e6}.

Fast nuclear reactors, like ADS or critical with Na cooling, feature a non-trivial neutron spectrum.
This allows for a lower production rate of radioactive waste and the burn-out of a fraction of the dangerous actinides.
Both the reactor dynamics
and burn-out of fuel and actinides depend on the neutron spectrum. Conventionally the neutron spectrum
in a reactor is measured by activation foil analysis.
Indeed, the activation of an isotope can be related to the convolution of the neutron flux
with the isotope activation cross section.
However, this complex, off-line, procedure
introduces large systematic uncertainties. A simple on-line technique is necessary for characterization
of reactor transients. For this purpose we developed a novel neutron spectrometer based on $^6$Li
converter sandwiched between two CVD diamond detectors. The energy of the incident neutron
converts completely into the energy of charged particles through the $^6$Li(n,$\alpha$)$t$ reaction.
This allows for event-by-event neutron energy measurement with the advantages of a solid state detector.
Moreover, the $^6$Li(n,$\alpha$)$t$ reaction is highly exothermic with $Q=4.7$ MeV, which
permits to reduce background by imposing a high detection threshold. The coincidence
between two crystals suppresses further noise and competing reactions.

In the present article we describe the experiment performed at TAPIRO,
which allowed, for the first time, a direct reconstruction of the fast reactor neutron spectrum.

\section{Spectrometer Description}
The spectrometer is based on the diamond detectors developed in Refs.~\cite{fulvio_r1,fulvio_r2} at the ``Tor Vergata''
University. These detectors are composed of three main layers: p-type diamond, intrinsic diamond and metal contact.
Such a scheme forms a diode-like structure. Once polarized in reverse bias mode by applying negative
voltage on the p-type (B-doped) cathode, a depletion layer is created in the intrinsic diamond.
The degenerate p-type layer acts as an ohmic contact.
Instead, the anode, creates a Schottky junction with the underlying intrinsic diamond.
In this configuration the electric current generated by the passage of an ionizing particle
in the depletion layer flows across the detector without any barrier. In fact, the Schottky
junction at the intrinsic diamond-metal interface accelerates electrons leaving the diamond bulk.

\begin{figure}[!t]
\centering
\includegraphics[bb=3cm 1cm 18cm 25cm, angle=270, width=3.5in]{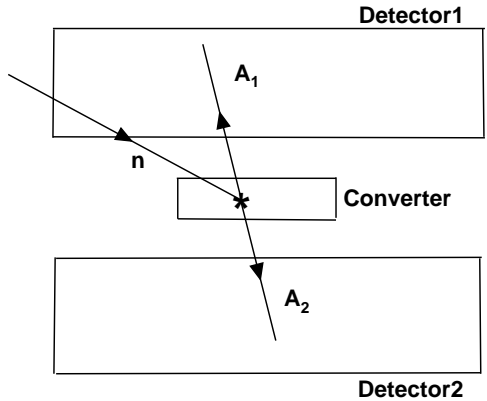}
\caption{Conceptual design of the spectrometer: the incident neutron is converted into two charged
particles inside the converter layer and these products are measured by two diamond detectors.}
\label{fig:sdw_scheme}
\end{figure}

The LiF compound was chosen as a neutron converter because of its chemical neutrality.
The selected LiF was enriched with $^6$Li isotope to 96\%.
The LiF layer is deposited on the top of metallic anode in such a way that $\alpha$ and $t$
can easily penetrate into the intrinsic diamond depletion layer with minimal loss of energy.
Then the two identical detectors, called SCD282 and SCD240, were placed with their LiF layers one in front of the
other leaving only 50 $\mu$m gap for the central ground microwire as shown in Fig.~\ref{fig:sdw_scheme}.

\section{TAPIRO Neutron Flux Measurements}
In order to prove capability of the spectrometer of measuring neutron spectrum in a fast reactor
the irradiation in TAPIRO was performed at ENEA of Casacia~\cite{enea_casaccia}.
TAPIRO is a fast source reactor with maximum power of 5 kW which features
an almost fission-like neutron spectrum with flux of $3\times 10^{12}$ n/cm$^2$s in its core.
The core made of 93.5\% enriched $^{235}$U (98.5\%) - Mo (1.5\%) alloy is shaped as a cylinder
with diameter of 12 cm and 12 cm tall. The reflector, divided in two parts,
inner - up to 18 cm and outer - up to 40 cm,
is made of electrolytic copper. This is followed by a biological shield, 1.75 m thick
borated concrete.
For the experiments the reactor features 3 irradiation channels in the median plane of the core,
radial channels 1 and 2 and diametral channel,
one tangential channel crossing the entire reactor 5 cm above it core
and the thermal column.
The mean neutron energy decreases rapidly with the distance from the core or the depth in the reflector,
allowing to select the desired spectrum by choosing the irradiation point.

\subsection{Experimental Setup}
\begin{figure}[!t]
\centering
\includegraphics[bb=0cm 0cm 25.5cm 25cm, width=3.5in]{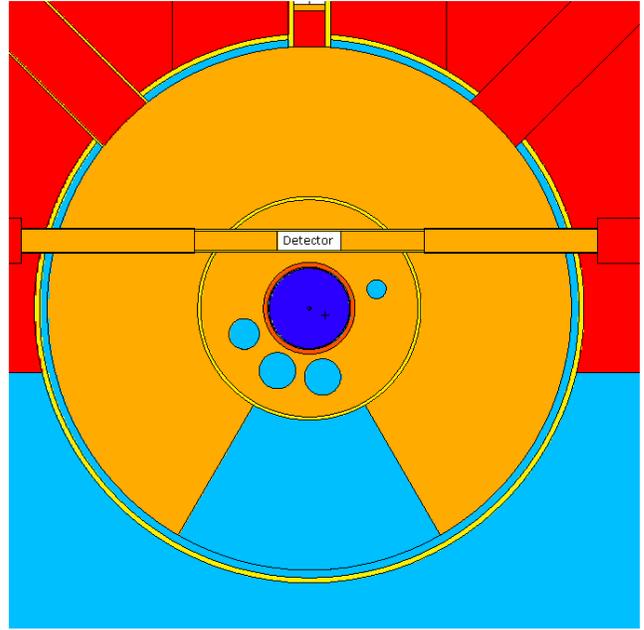}
\caption{Spectrometer location in the TAPIRO reactor (top view).
The dark blue color indicates the reactor core, yellow circles - inner and outer reflector,
long yellow rectangle - Tangential channel,
the detector is represented by the short white rectangle near the core.}
\label{fig:tapiro_fast_pos}
\end{figure}

\begin{figure}[!t]
\centering
\includegraphics[bb=0cm 6cm 20.5cm 24cm, width=3.5in]{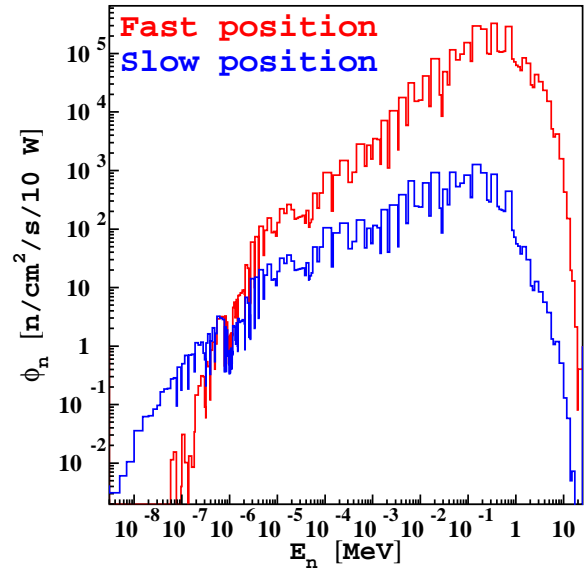}
\caption{Neutron fluxes in the two irradiation positions expected from MCNP simulations~\cite{alfonso_mcnp}
at 10 W reactor power.}
\label{fig:tapiro_expected_flux}
\end{figure}

The spectrometer was installed in the TAPIRO's tangential channel
at the closest point to the core (fast position). At this location the tangential channel has a diameter of 3 cm,
which gives the depth into the reactor inner reflector of 2.8-5.8 cm and
the radial distance from the core center of 10.6 cm.
This irradiation point is shown in Fig.~\ref{fig:tapiro_fast_pos}.
According to MCNP simulations~\cite{alfonso_mcnp,alfonso_mcnp2} shown in Fig.~\ref{fig:tapiro_expected_flux}
the mean neutron energy in this location is $<E_n>\simeq 0.586$ MeV
and the total neutron flux $\phi_n^{tot}\simeq 1.89\times 10^9$ n/cm$^2$/s/10 W.

For a better determination of resolution, for energy calibrations and for comparison between two neutron spectra
the spectrometer was also irradiated in the second (slow) position.
This second irradiation was performed with the tangential channel plug retracted by 30 cm
almost to the external edge of the outer reflector. This position was called slow
because of the significantly lower mean neutron energy $<E_n>\simeq 0.166$ MeV.
The total flux in this location is expected to be $\phi_n^{tot}\simeq 1.28\times 10^8$ n/cm$^2$/s/10 W.
The comparison of the fluxes in the two irradiation positions is shown in Fig.~\ref{fig:tapiro_expected_flux}.

The two sandwich spectrometer outputs were connected via a 5 m RG62 cable to custom fast charge amplifiers~\cite{cardarelli_amp}. At the amplifier inputs fast DC decouplers were
used to connect the detector bias voltage of -80 V/-120 V,
corresponding to 1.5-2 V/$\mu$m, supplied by an Ortec 710 module~\cite{ortec}.
The detector's signals boosted and shaped in the primary amplifiers
were further amplified by a Philips Scientific 771 amplifier~\cite{ps} with voltage gain set to $4$.
The overall gain of this chain was about 100 pVs/fC.
The amplified signals were sent to a SIS3305 10-bit digitizer~\cite{sis} working in 5 Gs/s mode.
The digitizer internal trigger was configured to fire on a coincidence~\cite{amp_preprint}
between the two spectrometer channels within a 64 ns interval.
The individual channel threshold was set to 40 mV.
The Data AcQuisition (DAQ) system was based on VME modular electronics. In particular,
a Concurrent Tech. VX813-09x single board computer~\cite{concur} was used as VME controller as well as
the acquisition host. The VX813-09x run 32-bit Centos 6 Linux operating system with
native Tsi148 VME controller drivers.
The SIS3305 was configured to generate interrupts on the VME bus when the number of events in its 2 Gb buffer exceeded
the imposed threshold.
When the interrupt was received by the VX813/09x controller all the data were copied
from SIS3305 buffer through a fast DMA MBLT transfer and sent to a secondary DAQ thread.
This secondary DAQ thread was buffering and saving the data on a fast Compact Flash card.
For every event $N_{samples}=960$ samples were saved in each channel for the total
waveform duration of 192 ns. Among these samples the signal length was $L_{signal}\sim 130$ samples or 25 ns.
Two TDC values of the common trigger as seen in the two channels were also saved.
The amount of data (about 240 kb/event) and event rate were relatively small,
thus VME transfer and disk writing speed were sufficient
to run the acquisition without additional delays.

During the experiment the reactor power was continuously monitored by a fission chamber.
Although the reactor power was fairly stable during irradiation the fission chamber current
was saved to the experiment log file for better accuracy.

\subsection{Data Analysis}
\begin{figure}[!t]
\centering
\includegraphics[bb=0cm 6cm 20.5cm 24cm, width=3.5in]{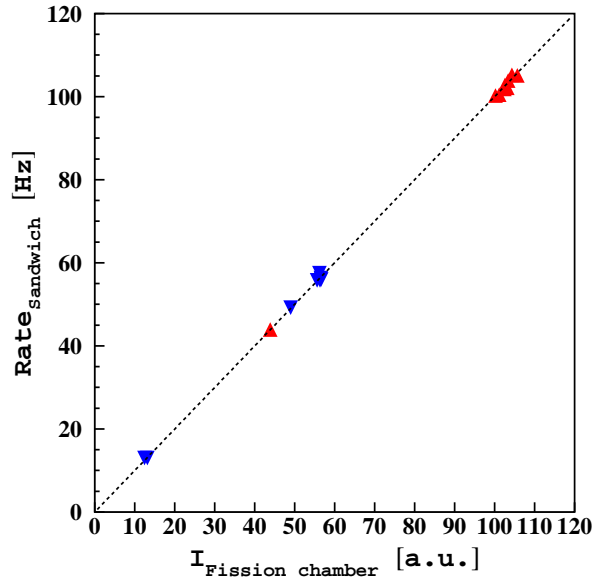}
\caption{Correlation between the spectrometer event rate and the current
of the power monitoring fission chamber. Measurements in different positions
are indicated by different colors.}
\label{fig:tapiro_fc_sdw_rates}
\end{figure}

For the absolute normalization of the measurement the TAPIRO power monitoring system combined with the
power-to-flux conversion from Refs.~\cite{pillon_tapiro_activation,alfonso_mcnp} were used.
The correlation between sandwich spectrometer event rate
and fission chamber current, shown in Fig.~\ref{fig:tapiro_fc_sdw_rates},
was found to be linear within uncertainties on the interval of one order of magnitude.
This normalization allows to correlate the measured absolute neutron flux
to a given reactor power and to compare this to the previous measurements~\cite{pillon_tapiro_activation}.

Although the coincidence trigger was fairly efficient in suppressing noise
the saved data were further filtered to discard remaining backgrounds.
This procedure included the following checks:
the event length (in samples) was checked to be the nominal one,
the integral of the first 100 samples before the signal were checked to be equal to the baseline within 3$\sigma$,
the trigger sample was checked to lay at the sample number $>50$ and $<N_{samples}-L_{signal}-50$,
the time of arrival of the signal from the two detectors was
requested to lay within 4.8 ns (1 ADC clock),
the integral of the signal peak had to be above the threshold in both channels.
All these conditions discarded $<1$\% of the recorded events.

The calibration of the energy deposited in each crystal was based on the energy
of the produced $t$, corrected for the energy lost in LiF layer and Cr anode.
The digitizer baseline was taken as the zero energy point.
The width of the $t$-peak was found to be $87$ keV for SCD282 and $104$ keV for SCD240.
These values were only slightly larger than the measured electronic's noise RMSs of $70$ and $77$ keV, respectively,
while the intrinsic broadening of the resolution due to the energy loss in LiF layer and Cr contacts was estimated
to be about $5$ keV.

The sum of the two energies deposited in the two crystals gives the incident neutron energy
plus $Q$-value of the reaction. The energy of thermal neutrons is smaller than
the energy resolution of the detector and therefore in this case the total measured energy is close
to the reaction $Q$-value. The spectrum of the sum of the energies of the two crystals builds as
an asymmetric peak at around 4.7 MeV. The peak width was found to be $160$ keV,
compared to the intrinsic resolution of $21$ keV due to energy losses in LiF layer and Cr contacts.

\begin{figure}[!t]
\centering
\includegraphics[bb=2cm 1cm 22cm 25cm, angle=270, width=3.5in]{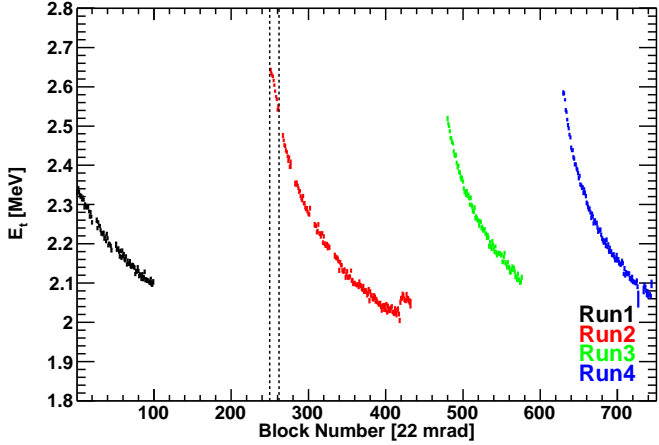}
\caption{Drift of the $t$ peak position with absorbed $\alpha+t$ dose. Detector bias supply was
turned off between the runs shown by different colors. Only the fraction of data selected by the two
dashed lines was kept in the present analysis.}
\label{fig:et_drift}
\end{figure}

\begin{figure}[!t]
\centering
\includegraphics[bb=2cm 1cm 22cm 25cm, angle=270, width=3.5in]{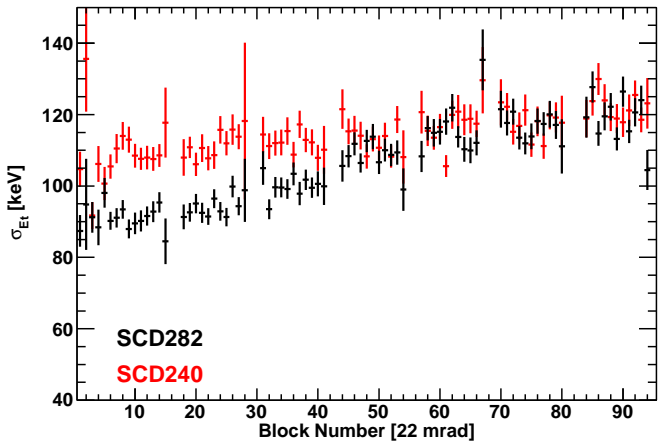}
\caption{Drift of the $t$ peak resolution with absorbed $\alpha+t$ dose for the two
diamond detectors of the spectrometer.}
\label{fig:sigma_et_drift}
\end{figure}

During the experiment it was noticed that the energy calibrations of two diamond detectors
changed with time as shown in Fig~.\ref{fig:et_drift}. This effect was attributed
to the build-up of space charge at one of the two electrical contacts of intrinsic diamond:
p-type (B-doped) layer or Cr anode.
The build-up of space charge near one contact screens the contact's potential,
thus reducing the electric field across the diamond depletion layer. This, in turn,
leads to a modification of the electrical signal shape, damping its height and
stretching its length.
Although the total collected charge remains the same a fraction of it could be cut off by
the finite integration window.
Choosing a longer signal integration time allows to compensate
this effect, but with the amplifier used in this experiment it leads to an unacceptable degradation of energy resolution.
Therefore, to correct for this drift of charge collection within fixed integration gate,
time-dependent calibration constants were applied.
The drift affected not only the energy calibration, but also the spectrometer
resolution. Indeed, a constant resolution in terms of signal charge after application
of time-dependent energy calibrations becomes time-dependent as well, as shown in Fig.~\ref{fig:sigma_et_drift}.
In order to maintain the best energy resolution only a small fraction, about 2\%, of the data
indicated by the two dashed lines in Fig.~\ref{fig:et_drift} was selected for this analysis.

\begin{figure}[!t]
\centering
\includegraphics[bb=2cm 1cm 22cm 25cm, angle=270, width=3.5in]{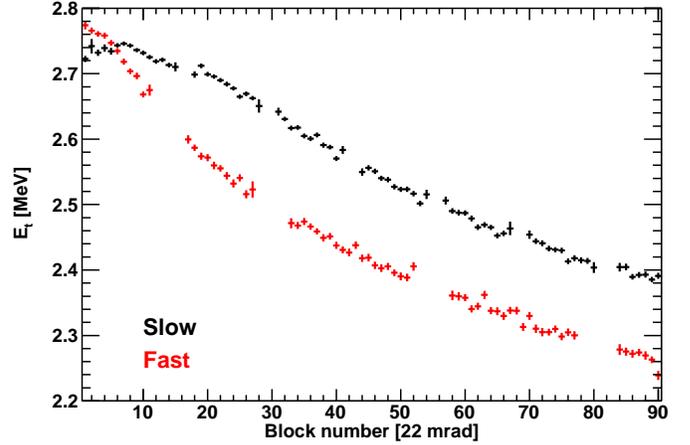}
\caption{Comparison of the drifts of the $t$ peak position in fast and slow neutron spectra.
The elastic scattering contribution is not included in the dose on $X$-axis.}
\label{fig:drift_cmp}
\end{figure}

It is worth noting that the drift of the observed charge collection is recoverable by turning off
the bias voltage. Indeed, in Fig.~\ref{fig:et_drift} the four different colors represent
the distributions taken in different runs, in between of which the bias voltage was off.
This observation confirms our interpretation of the drift as the space charge effect.
Another confirmation comes from the comparison of the drifts in slow and fast neutron fluxes
shown in Fig.~\ref{fig:drift_cmp}. The produced space charge depends on the dose deposited in the diamond
depletion layer. At the same dose produced by $\alpha$ and $t$ particles given in $X$-axis
the fast neutron spectrum generated a larger dose due to neutron elastic scattering off the carbon nuclei.
Thus in the fast neutron flux a more rapid drift was observed.

The absolute amount of $^6$Li deposited by evaporation on the diamond's Cr contacts was 
determined in a dedicated experiment~\cite{sdw_calib} with 20\% accuracy. This value
was used in the simulations to determine the correct absolute efficiency of the spectrometer.

The sandwich detector response to the neutron flux was modeled by the Geant Monte Carlo
library~\cite{geant4} version 4.9.5.p01.
The detector was described in details including:
PCB support, HPHT diamond substrate, B-doped CVD diamond cathode substrate, intrinsic CVD diamond,
Cr anode layer, LiF layer, Au deposited layers and Au coated copper wires.

The neutron flux of the TAPIRO fast reactor in the two irradiation positions taken from Ref.~\cite{alfonso_mcnp}
was generated isotropically on the spherical surface around the spectrometer of area about 1 cm$^2$.
The events with the energy deposited in both crystals of the sandwich sensitive volume
above the threshold were selected.

In order to normalize the simulations to the same neutron flux, seen by the spectrometer,
the reconstructed Geant4 distributions were rescaled by the factor:
\begin{equation}
L_{sim}= \phi_n t_{run} \frac{S_{gen}}{N_{gen}} ~,
\end{equation}
\noindent where $\phi_n$ is the neutron flux monitored by TAPIRO fission chamber,
$t_{run}$ is DAQ runtime, $N_{gen}$ is the total number of neutrons generated on the surface of area $S_{gen}$.

In order to compare Geant4 simulations to the data an additional Gaussian smearing due to
electronic readout noise was added to the reconstructed energies.
The obtained simulations describe the data fairly well, in particular in the region of $t$-peak,
which has almost a Gaussian shape. Instead the $\alpha$-peak is more asymmetric and features a larger
l.h.s. tail due to a higher energy loss of $\alpha$ particles.
The total deposited energy distribution is also well reproduced.

Similar simulations were performed with MCNP6~\cite{mcnp5}, but with Cr contacts of 100 nm
thickness which determined a larger energy loss of $\alpha$ and slight shift in its peak position.

\subsection{Results}
\begin{figure}[!t]
\centering
\includegraphics[bb=2cm 1cm 22cm 25cm, angle=270, width=3.5in]{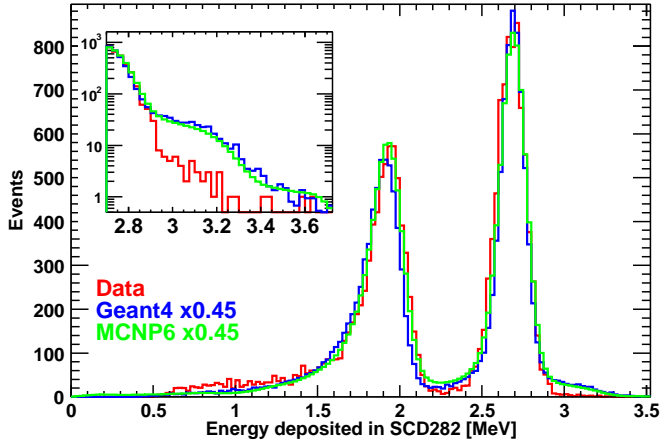}
\caption{Energy deposited in SCD282 of the spectrometer in comparison to Geant4 and MCNP6
simulations for the slow position. Because of the discrepancy in the absolute values between
expected and measured fluxes both simulations were rescaled by the factor 0.45.}
\label{fig:scd282_edep_th}
\end{figure}

The data measured in the slow neutron flux position were analyzed first in order to
calibrate better the spectrometer response in terms of deposited energy.
The energy deposited in a single spectrometer crystal was compared to Geant4
and MCNP6 simulations as shown in Fig.~\ref{fig:scd282_edep_th}.
Both simulations were obtained by generating neutron flux expected for
this location from Ref.~\cite{alfonso_mcnp} and normalized in absolute value
to the integrated flux of TAPIRO fission chamber. However, this comparison
revealed that the measured flux was a factor of two lower than expected.
Thus all the simulations were renormalized by the factor 0.45 for better qualitative comparison
with the data. Moreover, in the zoomed part of the spectrum around 2.9-3.3 MeV
a significant difference between data and rescaled simulations was found
suggesting further suppression of the measured flux in the neutron energy region 0.2-0.5 MeV
by a factor 4.

\begin{figure}[!t]
\centering
\includegraphics[bb=2cm 1cm 22cm 25cm, angle=270, width=3.5in]{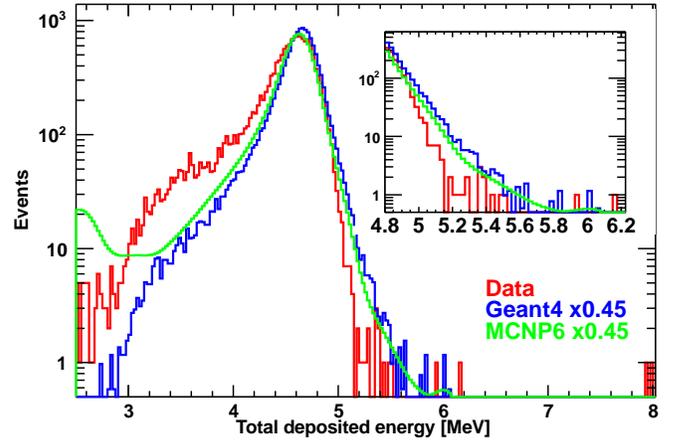}
\caption{Total energy deposited in the spectrometer in comparison to Geant4 and MCNP6
simulations for the slow position. Because of the discrepancy in the absolute values
between expected and measured fluxes both simulations were rescaled by the factor 0.45.}
\label{fig:tot_edep_th}
\end{figure}

The total energy deposited in the two crystals of the spectrometer, shown in Fig.~\ref{fig:tot_edep_th},
exhibits similar behavior. Also here both Geant4 and MCNP simulations were rescaled by the factor 0.45
for a better comparison. The high energy tail generated by neutrons with energies of 0.2-0.5 MeV,
corresponding here to 4.9-5.2 MeV deposited energy, is also strongly overestimated by the simulated flux.
Another difference, not so evident in the single crystal spectrum, is present
in the low energy tail. The ideal uniformity of Cr anode assumed in the simulations does not describe
its real distribution. Although the average Cr anode thickness is well reproduced,
as one can see from the comparison of the positions of the $\alpha$ peak in data and simulations,
small regions of thicker anode determine a larger $\alpha$ energy loss.
$t$ is less sensitive to this disuniformity because of its lower energy loss rate in matter.

\begin{figure}[!t]
\centering
\includegraphics[bb=2cm 1cm 22cm 25cm, angle=270, width=3.5in]{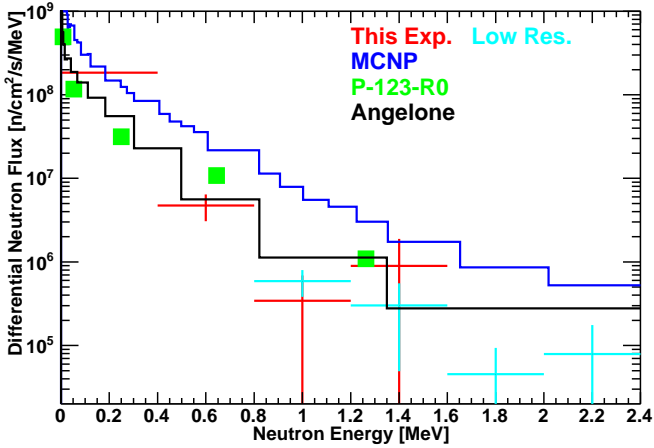}
\caption{Measured neutron flux in the slow position at 12 kW reactor power in comparison with MCNP simulations
from Ref.~\cite{alfonso_mcnp} (blue) and older evaluations from Ref.~\cite{enea_p-123-r0} (green)
and Ref.~\cite{angelone_flux_tapiro} (black). The latter distribution was rescaled
by the factor 0.17 due to position mismatch.
The cyan data points are from a low resolution but high statistics run.
The data error bars are statistical only, the systematic uncertainty consists of
and overall 20\% error due to $^6$Li quantity plus an energy dependent part due to
spectrometer resolution description.}
\label{fig:tapiro_th_flux}
\end{figure}

It is possible to extract the neutron flux using the distribution of the total deposited energy.
To this end the data and simulation distributions were rebinned in larger energy bins,
accounting for the spectrometer resolution. The achieved resolution of 160 keV on the total deposited
energy allowed to use 400 keV bins, which corresponded roughly to the FWHM.
The measured flux $\phi_n^{exp}$ in each energy bin was obtained as the ratio:
\begin{equation}
\frac{\partial \phi_n^{exp}}{\partial E_n}(E_n)= \frac{\partial \phi_n^{sim}}{\partial E_n}(E_n)
\frac{N_{data}(E_n)}{L_{sim} N_{sim}(E_n)} ~,
\end{equation}
\noindent where $\phi_n^{sim}$ is the simulated neutron flux,
$N_{data}(E_n)$ and $N_{sim}(E_n)$ are the numbers of data and simulation events in the given energy bin.
The obtained flux is shown in Fig.~\ref{fig:tapiro_th_flux} in comparison with MCNP simulations
from Ref.~\cite{alfonso_mcnp} (blue) and older evaluations from Ref.~\cite{enea_p-123-r0} (green)
and Ref.~\cite{angelone_flux_tapiro} (black). The data from Ref.~\cite{angelone_flux_tapiro}
were rescaled due to position mismatch by the factor 0.17, obtained as the ratio of MCNP
simulations in the two points.

\begin{figure}[!t]
\centering
\includegraphics[bb=2cm 1cm 22cm 25cm, angle=270, width=3.5in]{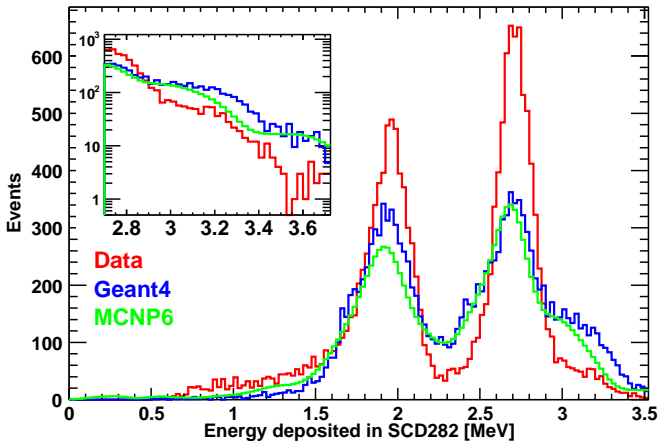}
\caption{Energy deposited in SCD282 of the spectrometer in comparison to Geant4 and MCNP6
simulations for fast position.}
\label{fig:scd282_edep}
\end{figure}

\begin{figure}[!t]
\centering
\includegraphics[bb=2cm 1cm 22cm 25cm, angle=270, width=3.5in]{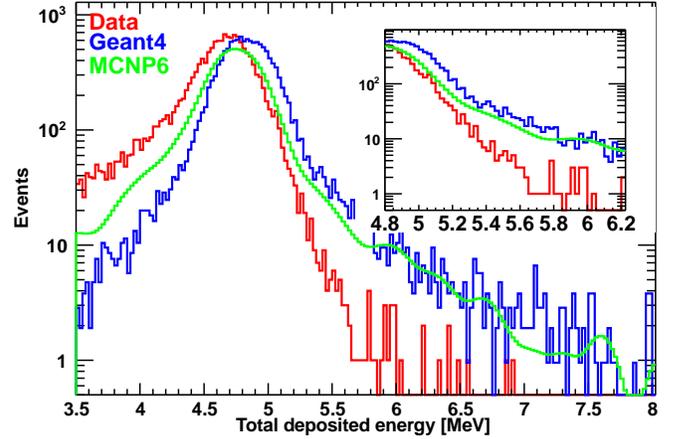}
\caption{Total energy deposited in the spectrometer in comparison to Geant4 and MCNP6
simulations for fast position.}
\label{fig:tot_edep}
\end{figure}

\begin{figure}[!t]
\centering
\includegraphics[bb=2cm 1cm 22cm 25cm, angle=270, width=3.5in]{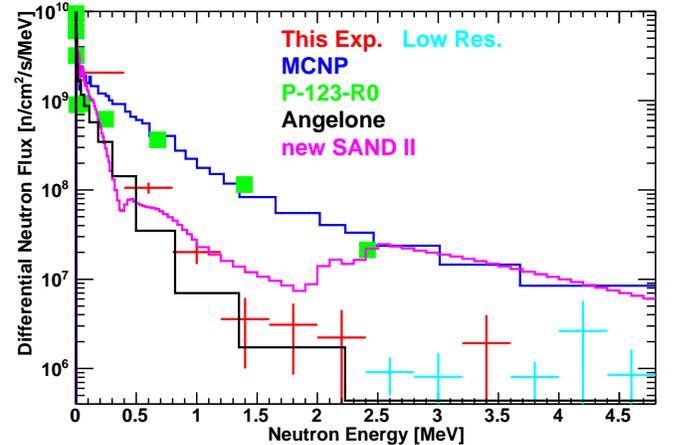}
\caption{Measured neutron flux in the fast position at 5 kW reactor power in comparison with MCNP simulations
from Ref.~\cite{alfonso_mcnp} normalized to activation foil analysis~\cite{pillon_tapiro_activation} (blue),
repeated SAND II activation analysis with modified input flux (magenta) and older evaluations from Ref.~\cite{enea_p-123-r0} (green)
and Ref.~\cite{angelone_flux_tapiro} (black). The latter distribution was rescaled
by the factor 2.56 due to position mismatch.
The cyan data points are from a low resolution but high statistics run.
The data error bars are statistical only, the systematic uncertainty consists of
and overall 20\% error due to $^6$Li quantity plus an energy dependent part due to
spectrometer resolution description.}
\label{fig:tapiro_flux}
\end{figure}

The same analysis was performed on the data taken in the fast neutron flux position.
These results are shown in Figs.~\ref{fig:scd282_edep}, \ref{fig:tot_edep} and \ref{fig:tapiro_flux}.
Qualitatively the fast position measurement showed larger disagreement
with the flux from Ref.~\cite{alfonso_mcnp}. Although the integrated neutron flux,
which comes mostly from the first energy bin was compatible within systematic uncertainties.
In fact, already the distribution of the energy deposited in a  single crystal,
shown in Fig.~\ref{fig:scd282_edep},
demonstrated a wrong ratio of fast and thermal neutrons.
In the total deposited energy distribution from Fig.~\ref{fig:tot_edep} the difference
resulted in an overall shift of the simulated distribution with respect to the measured one.

The reconstructed neutron flux, shown by red points in Fig.~\ref{fig:tapiro_flux},
is a factor of 5 lower at neutron energies above 0.4 MeV. This measured flux was used
to repeat the unfolding procedure~\cite{pillon_tapiro_activation} by means of SAND II~\cite{sand2} code.
The unfolding resulted in the flux shown in Fig.~\ref{fig:tapiro_flux} by magenta histogram,
which is in good agreement with the flux measured by our spectrometer up to 2 MeV (red points).
Above 2 MeV the data points from a low resolution but high statistics run were added.
These additional points disagree with activation analysis.

\section{Conclusions}
A novel neutron spectrometer for fast nuclear reactors based on $^6$Li converter sandwiched
between two CVD diamond detectors was used to measure neutron spectrum in the TAPIRO fast reactor.
The spectrometer was installed in the TAPIRO's tangential channel in two different positions:
fast position - 5 cm in the reflector and
slow position - at the edge of the reflector.
According to MCNP simulations~\cite{alfonso_mcnp} in these positions
the neutron fluxes of $\phi_n^{tot}\simeq 1.89\times 10^9$ n/cm$^2$/s/10 W
and $\phi_n^{tot}\simeq 1.28\times 10^8$ n/cm$^2$/s/10 W
with mean neutron energies $<E_n>\simeq 0.586$ MeV and $<E_n>\simeq 0.166$ MeV
were expected.
The measured fluxes were found to be different by a factor 0.45 at the slow position
and by a factor 0.9 at the fast position with 20\% overall systematic uncertainty.
Moreover, the neutron spectra in both positions resulted to be considerably softer,
such that $<E_n>^{exp}\sim 0.25$ MeV (fast) and $<E_n>^{exp}\sim 0.1$ MeV (slow).
The absolute normalization of MCNP simulations from Ref.~\cite{alfonso_mcnp}
was obtained from the activation foil irradiation at 3.5 kW reactor power
in the radial channel 1 at 5 cm in the reflector~\cite{pillon_tapiro_activation}.
Therefore, the expected fluxes come from an extrapolation of the measured spectrum
into different locations and to different reactor power. Although the distance from
the reactor core in Ref.~\cite{pillon_tapiro_activation} is similar to the current one
(fast position) the radial channel 1 is located at the middle
of the core, while the tangential channel passes above it. This leads to a harder
spectrum in Ref.~\cite{pillon_tapiro_activation}. This difference can be not completely
reproduced by MCNP model.
Moreover, high resolution data were
sensitive only to the neutron energy range 100 keV$<E_n<$2 MeV, while the activation foil analysis~\cite{pillon_tapiro_activation} did not have enough sensitivity in this interval.
Indeed, the 5\% sensitivity ranges given by SAND II unfolding code for the irradiated isotopes
excluded completely the region 230 keV$<E_n<$2.5 MeV. In order to verify this
statement the unfolding analysis was repeated with a new input flux
modified according to the values measured in this article. The analysis
converged to the solution compatible with our high resolution data.
The data measured with low resolution extend the covered energy range up to 4.8 MeV.
In this extended region 2.5 MeV$<E_n<$4.8 MeV our data are incompatible with Ref.~\cite{pillon_tapiro_activation}.
This could be explained also by the difference in reactor power, 700 times lower in the present experiment.

It must be emphasized that because of rapid fall of cross section with energy
the measured absolute flux in the first bin is highly dependent
on the initial neutron spectrum below 1 keV.
Because of the observed space charge build-up the analysis was performed only
on 2\% of the measured data featuring the best energy resolution.
With such statistics the neutron spectrum reconstruction was limited
to the region $E_n<2$ MeV.
The spectrometer energy resolution of 160 keV allowed for neutron spectrum
reconstruction in 400 keV bins. The resolution of a single crystal was two times
better and, in fact, it constrained much better the high energy part of the spectrum
by the comparison to simulations. However, the single crystal spectrum does not allow
to reconstruct neutron energy on a event-by-event basis.

The experiment had shown the following limitations of the present spectrometer prototype:
\begin{itemize}
\item the space charge build-up does not permit for a long continuous acquisition
and therefore limits the statistics,
\item relatively low energy resolution prevents from using narrow energy bins.
\end{itemize}
All these limitations will be addressed in the future prototypes.

\section*{Acknowledgment}
Authors would like to acknowledge excellent support provided during the experiment
by the TAPIRO facility staff and technical services of ENEA (Casaccia).
We also want to acknowledge the useful discussions with M.~Carta and
help with neutron flux simulations from A.~Santagata (ENEA, Casaccia).


\bibliographystyle{IEEEtran}
\bibliography{IEEEabrv,osipenko_39}
%

\end{document}